\documentclass{aa}
\def\folio{\ifnum\pageno=1\nopagenumbers\else\number\pageno\fi}

\def\lax    {\ifmmode{_<\atop^{\sim}}\else{${_<\atop^{\sim}}$}\fi}
\def\gax    {\ifmmode{_>\atop^{\sim}}\else{${_>\atop^{\sim}}$}\fi}
\newbox\grsign      \setbox\grsign=\hbox{$>$} 
\newdimen\grdimen   \grdimen=\ht\grsign
\newbox\simgreatbox \setbox\simgreatbox=\hbox{\raise.5ex\hbox{$>$}\llap
                        {\lower.5ex\hbox{$\sim$}}}\ht1=\grdimen\dp1=0pt
\newbox\simlessbox  \setbox\simlessbox =\hbox{\raise.5ex\hbox{$<$}\llap
                        {\lower.5ex\hbox{$\sim$}}}\ht2=\grdimen\dp2=0pt

\newbox\grsign \setbox\grsign=\hbox{$>$} \newdimen\grdimen \grdimen=\ht\grsign
\newbox\laxbox \newbox\gaxbox
\setbox\gaxbox=\hbox{\raise.5ex\hbox{$>$}\llap
     {\lower.5ex\hbox{$\sim$}}}\ht1=\grdimen\dp1=0pt
\setbox\laxbox=\hbox{\raise.5ex\hbox{$<$}\llap
     {\lower.5ex\hbox{$\sim$}}}\ht2=\grdimen\dp2=0pt
\def\gax{\mathrel{\copy\gaxbox}}
\def\lax{\mathrel{\copy\laxbox}}
\def\boxit#1    {\vbox{\hrule\hbox{\vrule\kern3pt
                  \vbox{\kern3pt#1\kern3pt}\kern3pt\vrule}\hrule}}
\def\h      {\ifmmode{^{\rm h}}\else{$^{\rm h}$}\fi}
\def\m      {\ifmmode{^{\rm m}}\else{$^{\rm m}$}\fi}
\def\s      {\ifmmode{^{\rm s}}\else{$^{\rm s}$}\fi}
\def\decas    {\ifmmode{{\rlap.}{''}}\else{${\rlap.}{''}$}\fi}
\def\mum     {\ifmmode{\mu{\rm m}}\else{$\mu{\rm m}$}\fi}
\def\s      {\ifmmode{^{\rm s}}\else{$^{\rm s}$}\fi}
\def\deg      {\ifmmode{^{\circ}}\else{$^{\circ}$}\fi}
\def\as     {\ifmmode {\rlap.}$\,$''$\,$\! \else ${\rlap.}$\,$''$\,$\!$\fi}
\def\decsec  {\ifmmode {\rlap.}$\,$^{s}$\,$\! \else ${\rlap.}$\,$^{s}$\,$\!$\fi}\def\decs  {\ifmmode {\rlap.}$\,$^{s}$\,$\! \else ${\rlap.}$\,$^{s}$\,$\!$\fi}

\def\kms    {\ifmmode{{\rm km~s}^{-1}}\else{km~s$^{-1}$}\fi}

\def\Mspy   {\ifmmode {M_{\odot} {\rm yr}^{-1}} \else $M_{\odot}$~yr$^{-1}$\fi}
\def\Mdot   {\ifmmode {\dot M} \else $\dot M$\fi}
\def\mhd    {\ifmmode {n_{{\rm H}_2}} \else $n_{{\rm H}_2}$\fi}
\def\mhcd   {\ifmmode {N_{{\rm H}_2}} \else $N_{{\rm H}_2}$\fi}

\def\El      {\ifmmode{E_{\ell}}\else{$E_{\ell}$}\fi}
\def\beam    {\ifmmode{\theta_{\rm B}}\else{$\theta_{\rm B}$}\fi}
\def\mjyb   {\ifmmode {{\rm mJy~beam}^{-1}} \else{mJy~beam$^{-1}$}\fi}
\def\mujyb   {\ifmmode {\mu{\rm Jy~beam}^{-1}} \else{$\mu$Jy~beam$^{-1}$}\fi}
\def\Trot   {\ifmmode{T_{\rm rot}}\else$T_{\rm rot}$\fi}    
    
\def\Teff   {\ifmmode{T_{\rm eff}}\else$T_{\rm eff}$\fi}

\def\ITRS   {\ifmmode{\smallint {\rm T}_{R}^{*}dv}\else{$\smallint 
{\rm T}_{R}^{*}dv$}\fi}
\def\ITRS   {\ifmmode{\smallint {\rm T}_{R}^{*}dv}\else{$\smallint 
{\rm T}_{R}^{*}dv$}\fi}
\def\ITAS   {\ifmmode{\smallint {\rm T}_{A}^{*}dv}\else{$\smallint 
{\rm T}_{A}^{*}dv$}\fi}

\def\lefttitle#1  {\noindent \hangindent=18.0pt \hangafter=1 {#1} \par}
\def\vol#1  {{\bf {#1}{\rm,}\ }}
\font\tenssb=cmssbx10
\font\tenbf=cmbx10
\font\sevenbf=cmbx8
\font\fivebf=cmbx6
\def\unetdemi    {\smallskipamount=6pt plus2pt minus2pt
                  \medskipamount=12pt plus4pt minus4pt
                  \bigskipamount=24pt plus8pt minus8pt
                  \normalbaselineskip=16pt plus0pt minus0pt
                  \normallineskip=2pt
                  \normallineskiplimit=0pt
                  \jot=6pt
                  {\def\smallskip {\vskip\smallskipamount}}
                  {\def\medskip   {\vskip\medskipamount}}
                  {\def\bigskip   {\vskip\bigskipamount}}
                  {\setbox\strutbox=\hbox{\vrule 
                    height17.0pt depth7.0pt width 0pt}}
                  \parskip 12.0pt
                  \normalbaselines}
\def\smallerspace {\smallskipamount=3pt plus0pt minus0pt
                  \medskipamount=6pt plus0pt minus0pt
                  \bigskipamount=10.5pt plus0pt minus0pt
                  \normalbaselineskip=10.5pt plus0pt minus0pt
                  \normallineskip=1pt
                  \normallineskiplimit=0pt
                  \jot=3pt
                  {\def\smallskip {\vskip\smallskipamount}}
                  {\def\medskip   {\vskip\medskipamount}}
                  {\def\bigskip   {\vskip\bigskipamount}}
                  {\setbox\strutbox=\hbox{\vrule 
                    height8.5pt depth3.5pt width 0pt}}
                  \parskip 0pt
                  \normalbaselines}
\def\memospace    {\smallskipamount=4pt plus1pt minus1pt
                  \medskipamount=6pt plus2pt minus2pt
                  \bigskipamount=14pt plus6pt minus6pt
                  \normalbaselineskip=14pt plus0pt minus0pt
                  \normallineskip=1pt
                  \normallineskiplimit=0pt
                  \jot=4pt
                  {\def\smallskip {\vskip\smallskipamount}}
                  {\def\medskip   {\vskip\medskipamount}}
                  {\def\bigskip   {\vskip\bigskipamount}}
                  {\setbox\strutbox=\hbox{\vrule 
                    height17.0pt depth7.0pt width 0pt}}
                  \parskip 2.0pt
                  \normalbaselines}
\def\memowidespace    {\smallskipamount=5pt plus1pt minus1pt
                  \medskipamount=7.5pt plus2pt minus2pt
                  \bigskipamount=17.5pt plus6pt minus6pt
                  \normalbaselineskip=17.0pt plus0pt minus0pt
                  \normallineskip=1.25pt
                  \normallineskiplimit=0pt
                  \jot=5pt
                  {\def\smallskip {\vskip\smallskipamount}}
                  {\def\medskip   {\vskip\medskipamount}}
                  {\def\bigskip   {\vskip\bigskipamount}}
                  {\setbox\strutbox=\hbox{\vrule 
                    height21.25pt depth8.75pt width 0pt}}
                  \parskip 2.5pt
                  \normalbaselines}
\usepackage{graphics,epsf}

\usepackage{natbib}
\usepackage{ulem}
      \def\new#1 {{\bf #1 }}
      \def\cut#1 {\sout{#1} }

\def\THCO {\hbox{$^{13}{\rm CO}$}}   %13CO

\def\TWCO {$\mathrm{^{12}CO}$} %12co
\def\THCO {$\mathrm{^{13}CO}$} %13co
\def\CSEO {$\mathrm{C^{17}O}$} %c17o
\def\percc {$\mathrm{cm^{-3}}$} %cm^-3
\begin{document}

\title{Mid- and high-$J$ CO observations towards ultracompact HII regions}
\author{F. Wyrowski, S. Heyminck, R. G\"usten, K.M. Menten}

%\\

\offprints{F. Wyrowski}

\institute{Max-Planck-Institut f\"ur Radioastronomie,
Auf dem H\"ugel 69, D-53121 Bonn, Germany
\email{wyrowski@mpifr-bonn.mpg.de}
}

\date{Received / Accepted}
\titlerunning{Mid- and high-$J$ CO observations towards UCHIIs}
\authorrunning{Wyrowski et al.}

% {Text of context}
% {Text of aims}
% {Text of methods}
% {Text of results}
% {Text of conclusions}

\abstract
{}
%aims
{A study of 12 ultracompact HII regions was conducted to probe the
  physical conditions and kinematics in the inner envelopes of the
  molecular clumps harboring them.}
%methods
{The APEX telescope was used to observe the sources in the CO (4--3) and
  \THCO\ (8--7) lines. Line intensities were modeled with the RATRAN radiative
  transfer code using power laws for the density and temperature to
  describe the physical structure of the clumps.}
%results
{All sources were detected in both lines. The optically thick CO (4--3)
 line shows predominantly blue skewed profiles reminiscent of infall.}
%conclusions
{Line intensities can be reproduced well using the physical structure
  of the clumps taken from the literature. The optically thick line
  profiles show that CO is a sensitive tracer of ongoing infall in
  the outer envelopes of clumps harboring ultracompact HII regions and
  hot molecular cores.}

%\keywords{ISM: molecules  -- Stars: circumstellar matter}


\maketitle

\section{\label{intro}Introduction}

%\begin{verbatim}
%
%high j obs in the past   
%     --> see PRE/high-j-co directory
%some details about 13co87: eupper, crit den
%
%\end{verbatim}

The earliest phases of massive star formation are still poorly
understood. We know that massive stars are being born in dense clumps
within giant molecular cloud complexes.
Ultracompact HII regions (UCHIIRs) embedded within these clumps
represent a key phase in the early
lives of massive stars (see review by Hoare et al.\ 2005).  UCHIIRs
were defined by Wood \& Churchwell (1989) to have sizes $\le 0.1$~pc,
densities $\ge 10^4$~\percc, and emission measures $\ge 10^7$~pc~$\rm
cm^{-6}$. In their environs, often hot ($T>100$~K), compact
($<0.1$~pc), and dense ($n({\rm H_2})>10^7$~\percc) cores are found, some of which are
believed to be in a stage prior to the formation of UCHIIRs (Kurtz et
al.\ 2000).

%Such
%cores with similar poperties are also observed without UCHIIRs within.
%These are
%believed to be in a stage (Kurtz et
%al.\ 2000).

A better understanding of these clumps %that harbor ongoing massive
%star formation 
is a crucial prerequisite for models of high-mass star
formation.  Several studies have attacked this topic in the past. In
particular, Hofner et al.\ (2000) conducted a \CSEO\ survey of 16
UCHIIRs and found typical sizes of 1~pc for the clumps and average
densities and temperatures of $10^5$~\percc\ and 25~K. These %average
values, obtained under simple assumptions, 
are an important first approximation but the high luminosities
and densities of the embedded hot cores show that a proper
representation of the physical conditions of the clumps requires
density and temperature gradients. Hatchell \& van der Tak (2003)
present models with power laws for the density and temperatures
that were constrained by the spectral energy distributions (SEDs) 
of the sources and single-dish
continuum images and CS line data. One drawback of using CS is the
abundance variation of this molecule with changing physical environments.
Toward the sources studied by Hatchell \& van der Tak (2003), the
abundance varies by almost three orders of magnitudes which suggests
that within an individual envelope the abundance might also vary
considerably. Therefore, in the inner regions of the clumps, mid- and
high-$J$ transitions of CO with a stable abundance in warm regions are
better probes of the physical conditions.

One of the important results in the study of low-mass star formation
has been the observation of infall motions (e.g. Belloche et al.
2002), which give direct evidence of accretion.
% processes of star formation. 
Toward
high-mass star-forming cores, the observational evidence of infall is
still very scarce (e.g. Wu \& Evans 2004). In addition to probing
density and temperature, mid- and high-$J$ lines of CO allow
the kinematics of the cores to be probed, hence to search for further evidence of
infall.

Here, we present a study of 12 UCHIIRs, mostly with associated hot
cores, in the CO (4--3) and \THCO\ (8--7) lines to probe the 
physical conditions
and kinematics of the inner envelopes of the clumps harboring the UCHIIRs.

\begin{figure*}
      \epsfysize=\textwidth 
      \rotatebox{-90}{\epsfbox{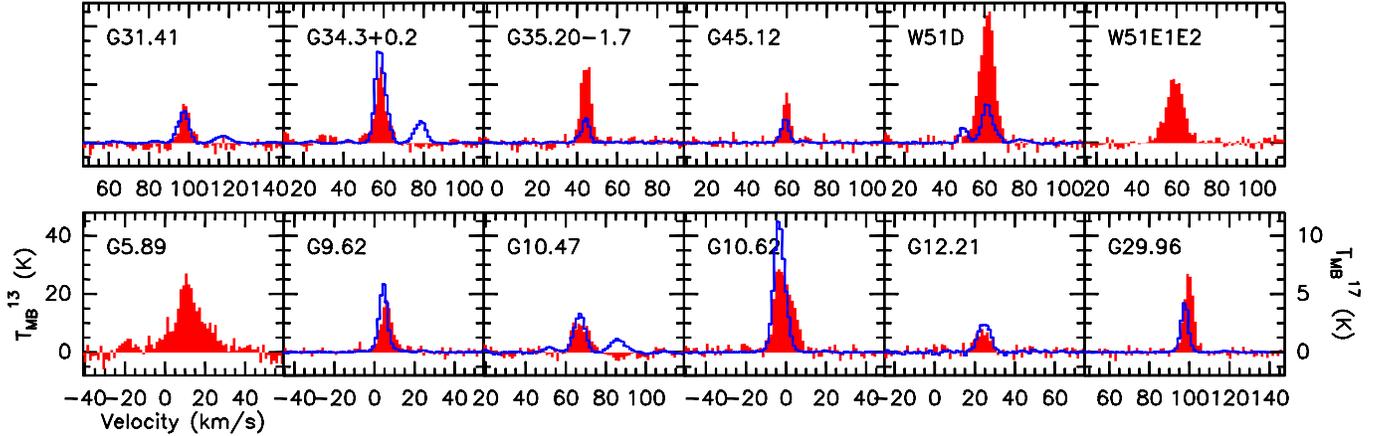}}
     \caption{ APEX \THCO\ (8--7) spectra (filled histograms) compared 
               with \CSEO\ (2--1) 
               observations (blue line, from Hofner et al.\ 2000).
               The spectral features appearing at --15 and 19~\kms\
                    in some of the \CSEO\ spectra are from other molecules.
                \label{fig:13co} }
   \end{figure*}

\section{\label{obs}Observations}

\begin{table}
\caption{Source list}
\begin{tabular}{lcc}
\hline
\hline
Source & R.A. (J2000) & Dec. (J2000) \\
\hline
G5.89--0.39  &  18:00:30.376  &  -24:04:00.48  \\
G9.62+0.19  &  18:06:15.000  &  -20:31:42.10  \\
G10.47+0.03  &  18:08:38.218  &  -19:51:49.71  \\
G10.62--0.38  &  18:10:28.661  &  -19:55:49.77  \\
G12.21--0.10  &  18:12:39.700  &  -18:24:20.00  \\
G29.96--0.02  &  18:46:03.950  &  -02:39:21.40  \\
G31.41+0.31  &  18:47:34.401  &  -01:12:45.95  \\
G34.26+0.15  &  18:53:18.499  &  01:14:58.66  \\
G35.20--1.74  &  19:01:46.440  &  01:13:23.50  \\
W51D  &  19:23:39.946  &  14:31:08.13  \\
W51E1E2  &  19:23:43.762  &  14:30:26.40  \\
G45.12+0.13  &  19:13:27.808  &  10:53:36.72  \\
\hline
\end{tabular}
\label{tab:sources}
\end{table}

%co(4-3): 13 sources (w/ n6334/g327)
%ci492:    8 sources (w/ n6334) 
%co(7-6)  11 sources
%13co(8-7): 12 sources

The observations were made with the Atacama Pathfinder Experiment
(APEX\footnote{This publication is based on data acquired with the
  Atacama Pathfinder Experiment (APEX). APEX is a collaboration
  between the Max-Planck-Institut f\"ur Radioastronomie, the European
  Southern Observatory, and the Onsala Space Observatory.}) in June
2005.  The frontend used was the MPIfR dual channel (460 and 810 GHz)
FLASH receiver (Heyminck et al.\ 2006, this volume), which enables simultaneous
observations of the \TWCO\ (4--3) and \THCO\ (8--7) lines. As
backends, the MPIfR Fast Fourier Transform Spectrometers (FFTS, Klein
et al.\ 2006, this volume) were used for the line observations.  The lines were
covered with 2048 channels within a bandwidth of 1 GHz, resulting in
spectral resolutions of 0.32 and 0.17~\kms. The spectra were converted
from $T_{\rm A}^*$ to $T_{\rm MB}$ units using forward efficiencies of 0.95
and beam efficiencies of 0.6 and 0.43 at 461 and 881 GHz, respectively.
The CO (4--3) line was observed for several sources on two different days and the
observed line temperatures agree within 20\%.
The beam sizes at the observing frequencies  are 13 and 7\arcsec.  All
observations were done in position switching mode on the UCHIIR
positions given in Table~\ref{tab:sources} using off-positions
250\arcsec\ to the east. These sources were selected from the sample
studied by Hofner et al.\ (2000) with the addition of the two sources
in W51. Pointing was done regularly on strong submm continuum sources in the
sample (G10.47 and G34.26) and should be accurate within 2\arcsec.

\section{\label{results}Results}

%\subsection{\THCO\ (8--7)}

\begin{figure*}
      \epsfysize=\textwidth 
      \rotatebox{-90}{\epsfbox{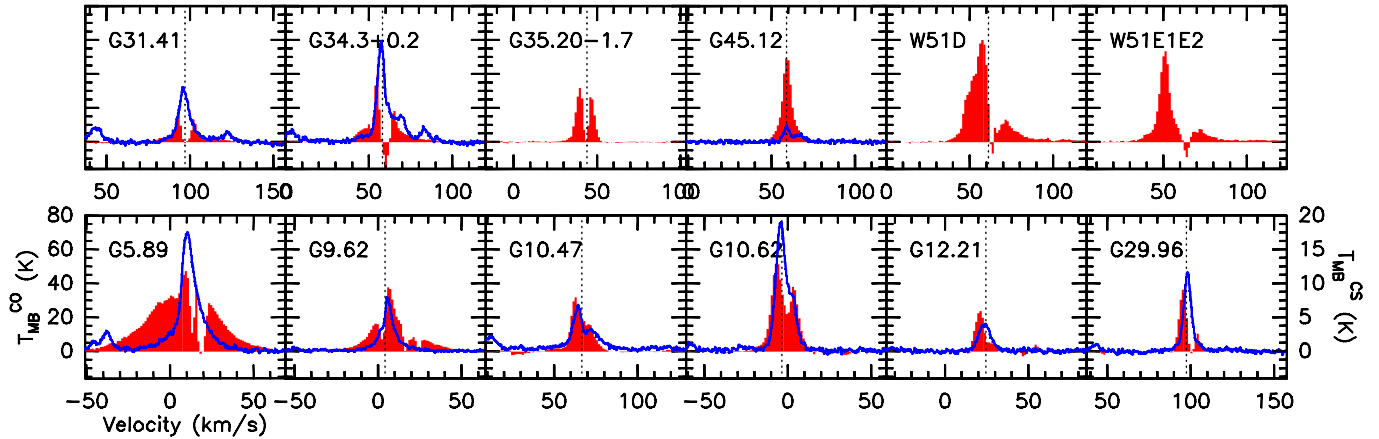}}
     \caption{ APEX CO (4--3) spectra (filled histograms) compared with CS (7--6)
               observations from Olmi \& Cesaroni (1999). The velocity from \CSEO\ is marked with 
               a dashed line. \label{fig:12co} }
   \end{figure*}

Figure~\ref{fig:13co} shows the observed \THCO (8--7) spectra compared
with \CSEO\ (2--1) spectra from Hofner et al.\ (2000) and line parameters from
Gaussian fits to the spectra are given in Table~\ref{tab:13co}. All
sources are clearly detected and show line shapes similar to \CSEO.
Notable deviation from Gaussian shapes are
seen in G10.62, with a strong red wing, and in G5.89--0.39, where the 
high velocity 
outflow wings are
prominent.  The \THCO\ linewidths are in all cases typically larger by
 2~\kms. While line broadening can be caused by high optical
depths, the absence of complicated self-absorbed spectral shapes
suggests rather moderate optical depths, hence a true increase in
motions. For most sources, the ratio of the \THCO (8--7) and \CSEO\ (2--1) 
line temperatures is the same
within a factor 2, with G35.20--1.7 and W51D being the exceptions with
much stronger \THCO\ lines.

\begin{table}
\caption{APEX \THCO\ (8--7) line parameters}
\begin{tabular}{lcccc}
\hline
\hline
Source & $T_{\rm MB}$ & $v_{\rm LSR}$ & FWHM \\
       &      (K)     & (km$\,$s$^{-1}$) & (km$\,$s$^{-1}$) \\
\hline
G5.89    & 20.8 ( 2.1 ) & 12.0 ( 0.4 ) & 20.9 (1.2 ) \\
G9.62    & 16.3 ( 1.0 ) & 6.2 ( 0.1 ) & 7.0 (0.3 ) \\
G10.47   & 11.5 ( 1.9 ) & 67.7 ( 0.5 ) & 12.4 (1.1 ) \\
G10.62   & 27.7 ( 1.3 ) & -1.1 ( 0.1 ) & 10.8 (0.3 ) \\
G12.21   & 6.9 ( 1.5 ) & 23.9 ( 0.5 ) & 9.7 (1.2 ) \\
G29.96   & 26.9 ( 1.0 ) & 99.9 ( 0.1 ) & 5.6 (0.2 ) \\
G31.41   & 14.0 ( 1.3 ) & 98.3 ( 0.2 ) & 6.6 (0.5 ) \\
G34.26   & 23.1 ( 2.1 ) & 58.8 ( 0.2 ) & 7.2 (0.6 ) \\
G35.20   & 27.3 ( 1.3 ) & 44.6 ( 0.1 ) & 6.0 (0.2 ) \\
W51D     & 43.9 ( 1.7 ) & 61.6 ( 0.1 ) & 8.7 (0.2 ) \\
W51E1E2  & 21.5 ( 1.3 ) & 59.4 ( 0.2 ) & 10.2 (0.4 ) \\
G45.12   & 17.3 ( 1.3 ) & 60.2 ( 0.1 ) & 4.1 (0.3 ) \\
\hline
\end{tabular}
\label{tab:13co}
\end{table}

%\subsection{\TWCO\ (4--3)}

The observed \TWCO (4--3) spectra are shown in Fig.~\ref{fig:12co}
compared with CS (7--6) spectra from Olmi et al.\ (1999). All sources
show complicated line profiles with self-absorption and outflow wings.
The self-absorption is in most cases redshifted compared to the
velocity of the optically thin \CSEO\ lines, with only G9.62 and 
G35.20 showing a blueshifted
self-absorption. The line shapes are either double peaked with a
stronger blue peak or skewed to the blue.  G9.62 is again an exception
and G45.12 shows a rather symmetric profile. G5.89 is dominated by the
strong outflow and absorbing foreground clouds (Klaassen et al.\
2006). Also in the red line wing of G9.62, absorption due to
foreground clouds is seen, consistent with the HCO$^+$ observations
by Hofner et al.\ (2001). The asymmetries in the line shapes of the CS (7--6) 
line are
mostly consistent with those we find in \TWCO (4--3) but the self
absorption is weaker. To quantify the asymmetries seen in the profiles
to check for infall signatures (Table~\ref{tab:12co}), we determined
the ratio of blue-to-red peak intensity for the 8 sources with 2 peaks. Seven
sources show significantly stronger blue peaks, and the average ratio
of the sample is 3.3, thus indicative of infall. For the remaining 3
sources with only one peak (G5.89 was omitted due to its strong
outflow), we 
determined $ \delta v = (v_{\rm thick}-v_{\rm thin})/\Delta v_{\rm thin}$, 
the difference between optically
thick (\TWCO) and thin (\CSEO) line peaks over the optically thin line widths, which
can be used as an infall indicator (Mardones et al.\ 1997). For 2
sources $\delta v$ is $-0.5$, hence clearly blue-shifted, and for 1 it
is close to 0.

\begin{table}
\caption{APEX \TWCO\ (4--3) collapse indicators. For two-peaked
         profiles the ratio of blue and red peak and
         for the single-peaked lines the skewness parameter $ \delta v$ are given.
          The profile
         column denotes blue (B) and red (R) line profiles. }
\begin{tabular}{lcccc}
\hline
\hline
Source   & $T_{\rm blue}/T_{\rm red}$ & $\delta v$ & Profile \\
%         &                         &            &         \\
\hline
%%G5.89    &       & 12.0 ( 0.4 )      & 20.9 (1.2 ) \\
G9.62    & 0.44  &       & R \\
G10.47   &       & -0.50 & B \\
G10.62   & 1.38  &       & B \\
G12.21   &       & -0.51 & B \\
G29.96   & 4.21  &       & B \\
G31.41   & 1.40  &       & B \\
G34.26   & 2.03  &       & B \\
G35.20   & 1.42  &       & B \\
W51D     & 4.67  &       & B \\
W51E1E2  & 7.87  &       & B \\
G45.12   &       &  0.04 &   \\
\hline
\end{tabular}
\label{tab:12co}
\end{table}

\section{\label{discussion}Discussion}

\begin{figure}
\begin{center}
%%\figurenum{2}
%      \epsfysize=\columnwidth
      \epsfysize=7cm
      \rotatebox{-90}{\epsfbox{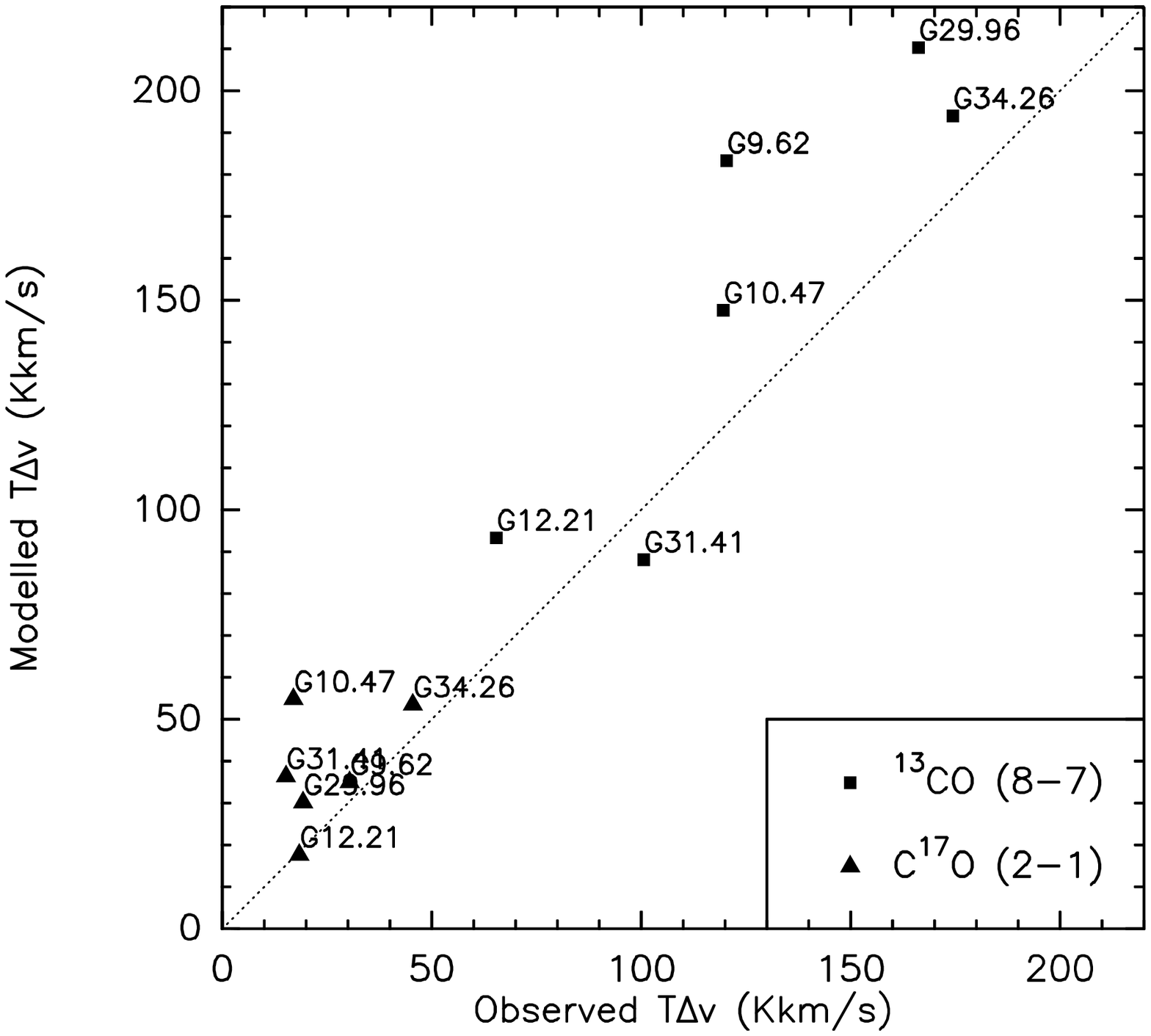}}
     \caption{ Comparison between modeled and observed integrated 
               \CSEO\ and \THCO\ intensities. The modeled intensities 
               were obtained using the physical structure of the clumps determined
               by Hatchell \& van der Tak (2003).
               \label{fig:compare} }
\end{center}
   \end{figure}

The CO emission of the sources has been modeled with the Monte Carlo
radiative transfer program of Hogerheijde \& van der Tak (2000).
Besides collisional excitation, dust radiation is taken into account
using grain properties from Ossenkopf \& Henning (1994), Model 5. The
radial profile of the density, as well as inner and outer radii, were
taken from the best-fit DUSTY models of Hatchell \& van der Tak (2003,
RATRAN code) based on single-dish maps of dust continuum and CS
molecular line emission. They determined power laws for the density
structure with power law exponents between -1.5 and -2.0 for our
sample. RATRAN and DUSTY were used with the same dust properties. The
temperature structure was then solved by DUSTY (Ivezi{\' c} et al.
1997) in a self-consistent way.  An isotopic ratio of 40 and 2000 was
used for \THCO\ and \CSEO, respectively, relative to \TWCO. 
To simplify
the modeling, no variable velocity field was used but only constant, 
turbulent line
widths. This is a reasonable approximation of the integrated
intensities of lines that are optically thin or have only moderate
optical depths; but for optically thick lines like the \TWCO\ (4--3)
line, the resulting integrated intensities will depend crucially on the
assumed velocity fields. Therefore, in Fig.~\ref{fig:compare} only the
integrated \THCO\ (8--7) and \CSEO\ (2--1) line intensities from the
observations and the modeling are compared. 
For this comparison, the modeled intensities were determined from
the RATRAN output images by convolving with the APEX and 30~m observing beams.
 The agreement between the
modeled intensities using the Hatchell \& van der Tak fit results and
the observations is remarkable.  The only large deviations are the
predictions of the \CSEO\ intensities for G10.47 and G31.41 for which
the outer radius of the clumps might be overestimated, and this mostly affects
the \CSEO\ (2--1) lines and not \THCO\ (8--7).  Since the
modeled lines probe very different excitation regimes and are
completely independent of the data used by Hatchell \& van der Tak,
this agreement is a strong validation of the power law structure in
$T$ and $n$ of the clumps.

The observed \TWCO\ profiles are very reminiscent of infall
asymmetries routinely detected toward low-mass star-forming cores
(e.g. Myers et al.\ 2000). For individual sources, there is still the
possibility that some of the absorption might be caused by colder
foreground clouds, but this would not explain the general trend of
redshifted absorption and most of the blue peaks being stronger.
Also, in several cases the observed infall signature is consistent
with results from interferometric observations where the infall is
observed mostly in absorption using different molecules. Examples are
G10.62 (NH$_3$, Keto et al.\ 1988), W51E (HCO$^+$, Rudolph et al.\
1990) and G29.96 (HCO$^+$, Maxia et al.\ 2001). We can infer a total of 9 out of 11
sources with infall evidence from the line profiles
discussed in Sect.~\ref{results}.
 The excess parameter $E=(N_{\rm blue}-N_{\rm red})/N_{\rm tot}$, introduced by
Mardones et al.\ (1997) to quantify the statistics of infall surveys,  is
for our sample 0.7, which is larger
than for low-mass samples or the massive star-forming region survey by Wu
\& Evans (2003).  This might be a selection effect, since mostly
UCHIIRs with associated hot cores, hence early stages of massive star
formation, were targeted.  It might also be related to using CO as an
infall tracer. The excess parameter using CS for our sample would have
been smaller. The high optical depths of the CO (4--3) line make it
a very sensitive probe of infalling motions in the outer envelopes
of the massive star-forming clumps. Therefore, while observations
of much higher-$J$ optically thick CO lines are needed to probe
the kinematics in the inner part of the clumps, the CO (4--3) observations
clearly show that in the outer parts (a remnant) infall is still going on. 

In the near future, the upcoming CHAMP+ array -- a 2x7 pixel
heterodyne array for simultaneous observations in the 350 and
450~$\mu$m atmospheric windows to be installed at the APEX telescope
in the middle of 2006 -- will allow us to image high-$J$ CO lines and then,
together with proper models of the density and temperature structure,
put further and tighter constraints on the velocity structure of the
massive star-forming clumps.

%the 12co lines show too much self absorption and of course cannot
%reproduce the observed asymmetries. These asymmetries of the very
%optical thick 12co lines can better fit with models dominated by
%velocity gradients but small turbulent widths (see cesaroni 1995).

% \begin{verbatim}
%
%           for selected sources compute 12co with velocity field:
%
%                    v(r) and dv(r)
%
%           --> compute spectra with RATRAN
%
% give outlook of champ+ potential to constrain models
%
% infall: compare with Wu&Evans (g9 expansion also seen in hcn)
%
% if modeling of infall fails, just qualitative analysis as in wu&evans
%
% \end{verbatim}

We thank Riccardo Cesaroni for providing the CS (7--6) spectra in
   CLASS format.

%\begin{figure}
%\figurenum{2}
%      \epsfxsize=7cm 
%      \rotatebox{-90}{\epsfbox{13co-c17o.eps}}
%
%     \caption{ line ratio \label{fig:13co-c17o} }
%
%   \end{figure}

%\begin{figure}
%%%\figurenum{2}
%      \epsfysize=7cm 
%      \rotatebox{-90}{\epsfbox{grid.eps}}
%%
%     \caption{ Results of LVG modelling \label{fig:grid} }
%
%   \end{figure}

%\begin{figure}
%\figurenum{2}
%      \epsfysize=8cm 
%     \rotatebox{-90}{\epsfbox{g1062-compare.eps}}
%
%     \caption{ THIS FIGURE WILL BE REPLACED WITH A COMPARISON
%               OF AN INFALL PROFILE WITH MODEL RESULTS ALTHOUGH
%               THERE MIGHT NOT MUCH SPACE LEFT FOR A DISCUSSION OF IT.
%               \label{fig:g1062} }
%
%   \end{figure}

\end{document}